\documentclass[reprint,aps,prl,showpacs,footnoteinbib,amsmath,amssymb]{revtex4-1}
\pdfoutput=1 
\usepackage{amssymb}
\usepackage{amsmath}
\usepackage{graphicx}

\newcommand{\bea}{\begin{eqnarray}}
\newcommand{\eea}{\end{eqnarray}}

\begin{document}
	
\title{Fermi polaron-polaritons in charge-tunable atomically thin semiconductors}
\author{Meinrad Sidler}
\author{Patrick Back}
\author{Ovidiu Cotlet}
\author{Ajit Srivastava$^{\dagger}$}
\author{Thomas Fink}
\author{Martin Kroner}
\author{Eugene Demler$^*$}
\author{Atac Imamo\u{g}lu}

\affiliation{Institute of Quantum Electronics, ETH Z\"{u}rich, CH-8093
Z\"{u}rich, Switzerland.}

\affiliation{$^{\dagger}$Physics Department, Emory University, Atlanta, Georgia 22138, USA}

\affiliation{$^*$Physics Department, Harvard University, Cambridge, Massachusetts 02138, USA}

\date{\today }

\begin{abstract}
The dynamics of a mobile quantum impurity in a degenerate Fermi
system is a fundamental problem in many-body physics. The interest
in this field has been renewed due to recent ground-breaking
experiments with ultra-cold Fermi gases
~\cite{massignan2014polarons,zwierleinPRL2009,schmidtPRA2012,kohstall2012metastability,koschorreck2012attractive}.
Optical creation of an exciton or a polariton in a two-dimensional
electron system embedded in a microcavity constitutes a new frontier
for this field due to an interplay between cavity-coupling favoring
ultra-low mass polariton formation~\cite{carusotto2013quantum} and
exciton-electron interactions leading to polaron or trion
formation~\cite{esser2001theory,ganchev2015three}. Here, we present
cavity spectroscopy of gate-tunable monolayer
MoSe$_2$~\cite{xu2013TMDreview} exhibiting strongly bound trion and
polaron resonances, as well as non-perturbative coupling to a single
microcavity mode~\cite{liu2015strongcouplingTMD,dufferwiel2015}. As
the electron density is increased, the oscillator strength
determined from the polariton splitting is gradually transferred
from the higher-energy repulsive-exciton-polaron resonance to the
lower-energy attractive-polaron manifold. Simultaneous observation
of polariton formation in both attractive and repulsive branches
indicate a new regime of polaron physics where the polariton
impurity mass is much smaller than that of the electrons.  Our
findings shed new light on optical response of semiconductors in the
presence of free carriers by identifying the Fermi polaron nature of
excitonic resonances and constitute a first step in investigation of
a new class of degenerate Bose-Fermi
mixtures~\cite{laussy2010exciton,cotlet2015superconductivity}.
\end{abstract}

\maketitle

\paragraph*{} 
Transition metal dichalcogenide (TMD) monolayers represent a new
class of two dimensional (2D) semiconductors exhibiting features such as strong
Coulomb interactions~\cite{heinzPRL2014}, locking of spin and valley
degrees of freedom due to large spin-orbit
coupling~\cite{xu2013TMDreview} and finite electron/exciton Berry
curvature with novel transport and optical
signatures~\cite{ajitPRL2015,xiaoPRL2015}. Unlike quantum wells or
two-dimensional electron systems (2DES) in III-V semiconductors, TMD
monolayers exhibit an ultra-large exciton binding energy $E_{exc}$ of order
$0.5$~eV~\cite{heinzPRL2014} and strong trion peaks in
photoluminescence (PL) that are redshifted from the exciton line by
$E_T \sim 30$~meV~\cite{xu2013TMDreview,urbaszek2015}. These features provide a
unique opportunity to investigate many-body physics associated with
trion~\cite{berkelbachPRB2013} formation as well as coupling of
excitons to a 2DES~\cite{heinzPRL2015} and to cavity
photons~\cite{rapaportPRB2001,smolka2014cavity}, provided that the
experimental set-up allows for varying the electron density $n_e$ and
light matter coupling strength $g_c$.

Here, we carry out an investigation of Fermi
polarons~\cite{massignan2014polarons} in a charge-tunable MoSe$_2$
monolayer embedded in an open microcavity structure (Fig.~1a-b).
Since $E_{exc}$ is much larger than all other relevant energy
scales, such as the normal mode splitting ($2 g_c$), $E_T$ and the
Fermi energy ($E_F$), an optically generated exciton in a TMD
monolayer can be considered as a robust mobile bosonic impurity
embedded in a fermionic reservoir (Fig.~1c). The Hamiltonian
describing the system is
\begin{eqnarray}
\label{eq1}
H &=&   \omega_c c_0^{\dagger} c_0 + \sum_k \omega_X (k) x_k^{\dagger} x_k +   g_c (c_0^{\dagger} x_0 + h.c.) \nonumber \\
 & +&  \sum_k \epsilon_k e_k^{\dagger} e_k + \sum_{k,k',q} V_q ( x_{k+q}^{\dagger}e_{k'-q}^\dagger e_{k'} x_{k} + h.c.) \;,
\end{eqnarray}
where the first line describes the coupling of 2D excitons,
described by the exciton annihilation operator $x_k$ and  dispersion
$\omega_X (k)= -E_{exc} + k^2/2m_{exc}$, to a zero-dimensional (0D)
cavity mode $c_0$ whose resonance frequency $\omega_c$ can be tuned
by applying a voltage ($u_p$) to a piezoelectric actuator that
changes the cavity length. This part of the Hamiltonian corresponds
to the elementary building block of the recent ground-breaking
experiments based on coupled 0D-polariton systems \cite{amo2014}.
The second line of the Hamiltonian describes the Feshbach-like
physics associated with the bound-molecular (trion) channel and the
corresponding effective interactions between the excitons and the
electrons~\cite{massignan2014polarons}. Provided that the
exciton-electron coupling can be treated as  an attractive contact
interaction,  $E_T$ directly determines the interaction strength
$V_q = V_0$~\cite{parish2011polaron}. The Hamiltonian of
Eq.~(\ref{eq1}) therefore combines the physics of cavity-polaritons
with that of Fermi polarons.

The Fermi polaron problem describes the screening of a mobile
impurity via generation of  particle-hole pairs across the Fermi
surface (Fig.~1c). When the impurity is a fermion with different
spin, this problem corresponds to the highly polarized limit of a
strongly interacting Fermi system~\cite{massignan2014polarons}. The
corresponding systems exhibit a wealth of complex phenomena, such as
the elusive Fulde-Ferrell-Larkin-Ovchinikov pairing mechanism, the
Chandrasekar-Clogston limit of BCS superconductivity and itinerant
ferromagnetism \cite{duine2005itinerant}. While the cavity
spectroscopy we implement to study the exciton-2DES problem in the
weak coupling regime is analogous to the rf spectroscopy of
impurities in degenerate Fermi gases~\cite{zwierleinPRL2009}, the
strong coupling regime of the TMD-monolayer-microcavity system
represents a new frontier for quantum impurity physics. More
specifically, since the exciton-polariton dispersion can be tuned by
changing $\omega_c(u_p) - E_{exc}$ to yield an effective polariton
mass that is (up to) four orders of magnitude smaller than that of
the electron~\cite{carusotto2013quantum}, the extension of our
experiments to a 2D cavity could realize a Fermi-polaron system with
a tunable ultra-small mass impurity. We find that a theoretical
model based on a truncated basis approach (Chevy
ansatz)~\cite{parish2013highly} treating the system as an excitonic
Fermi polaron captures the experimental signatures such as the
$n_e$-dependent blueshift of the exciton resonance and the
oscillator strength transfer from the repulsive exciton polaron to
the attractive polaron~\cite{esser2001theory}.

\paragraph*{} 
We embed a MoSe$_2$/hBN/graphene
heterostructure~\cite{xuNatComm2013} inside an open optical cavity~\cite{besga2015}
consisting of a flat dielectric mirror and a fiber-mirror with a
radius of curvature of $30 \,\mu$m (Fig.~1a). We use the graphene
layer as a top gate that controls the electron density in the
MoSe$_2$ layer (Fig.~1b), allowing us to tune the Fermi energy from
$E_F = 0$ to $E_F \ge E_T$. The thickness of the hBN layer is
chosen to ensure that the MoSe$_2$ is located at an anti-node of the
cavity, while the graphene monolayer is at a node where the
intra-cavity field vanishes; this choice ensures that the graphene
absorption does not lead to a deterioration of the cavity finesse
which we estimate to be ${\cal F} \sim 200$.

In order to characterize the elementary optical excitations of the
MoSe$_2$ monolayer, we set the cavity length to $L_{cav}= 9.1 \,\mu$m
and carry out spectroscopy  in the limit of weak (perturbative)
coupling to the cavity mode. Figure~2a depicts the cavity
transmission spectrum in this weak coupling regime obtained for $V_g
= -3$~V: the parallel diagonal lines correspond to transmission
maxima associated with neighboring axial modes of the cavity.
Zooming in to the central mode, we find that the mode energy as well
as its linewidth varies non-trivially due to coupling to the
MoSe$_2$ excitations. We plot the color-coded cavity line broadening
(Fig.~2b) and line-shift (Fig.~2c) as a function of $V_g$ (vertical
axis) and the fundamental cavity mode frequency (horizontal axis).
Since the bare cavity linewidth ($\sim 0.38$~meV) is much smaller
than the spectral features associated with exciton-polaron and trion
resonances, the increase in cavity linewidth, or shift in cavity
resonance frequency allows us to determine the imaginary
(absorption) and real (dispersion) parts of MoSe$_2$ linear
susceptibility (Fig.~2b-c). We note that even for $L_{cav}= 9.1
\,\mu$m, the exciton resonance is in the strong coupling regime for
$V_g < -10$V, albeit with a small normal mode splitting: in this
limit, highlighted by the dashed rectangle in Fig.~2b, it is not
possible to directly measure the cavity line broadening (see
Supplementary Information). As a finite electron density $n_e$ is
introduced by increasing $V_g$ to $-10$~V, a new absorption
resonance (shaded blue curve), which is red-detuned by $\sim 25$~meV
from the bare exciton resonance, emerges (Fig.~2d). At the same
time, the exciton line blueshifts and broadens, thereby ensuring
that the coupling to the cavity mode is in the perturbative limit.
For $V_g = 0$~V, the exciton resonance sharply shifts to higher
energies as the lower-energy resonance becomes prominent (Fig.~2e).
Further increase in $V_g$ leads to an increasing energy of the
redshifted resonance and an indiscernible exciton feature (Fig.~2f).
We observe that for $V_g > 20$~V, the MoSe$_2$ monolayer-induced
cavity line broadening exhibits a spectrally flat blue tail in
absorption (Fig.~2g). Since the cavity line broadening (Fig.~2b) and
line-shift (Fig.~2c) data are connected by Kramers-Kronig relations,
we mainly refer to line broadening in the following discussion.

We note that TMD monolayer PL at low $n_e$ is known to exhibit sharp trion peaks~\cite{xuNatComm2013}: photo-excited carriers predominantly relax to the lowest energy molecular (trion) state,
which in turn decays by spontaneous emission to an excited state of the 2DES. We observe that in
Figure~2e the PL line (green curve) is only slightly redshifted with
respect to the peak in absorption. In contrast, increasing $V_g$ further
results in a redshift of the PL peak while the low-energy absorption
peak experiences a blueshift (Fig.~2f). Further increase in $n_e$ results in a large splitting exceeding $40$~meV between the absorption and PL peaks (Fig.~2g), suggesting that they are associated with different elementary optical excitations.

We identify the emerging lower-energy resonance in absorption for $V_g
\ge -10$~V as stemming from attractive-exciton-polarons~(Fig.~1c). The
observation of substantial line broadening of the cavity mode
indicate a sizable overlap between the ground state (with no
optical or electronic excitation above the Fermi sea) and the
optically excited state. These observations in turn render it
unlikely that the observed features are associated with direct
optical excitation of a trion. In contrast to the latter, an attractive polaron has a finite amplitude for having no electron-hole pair excitation across the Fermi surface ensuring a sizable quasi-particle weight. The strong $n_e$-dependent blueshift of the exciton resonance in turn indicates that it should be identified as the repulsive polaron -- a metastable excitation of the interacting electron-2DES system~\cite{schmidtPRA2012}.

The most spectacular signature demonstrating the polaron nature of
the  absorption resonances is obtained in the strong coupling regime
of the interacting cavity-exciton-electron system, which is reached
by decreasing the effective cavity length to $\sim 1.9\, \mu$m.
Figure~3a-c shows the transmission spectrum as the cavity length is
changed by $\sim 100$~nm at three different gate voltages. The
observation of normal mode splitting when the cavity mode is tuned
into resonance with the lower-energy absorption resonance
(Fig.~3b-c) demonstrates the large overlap between the initial and
final states of this optical transition which in turn proves that
the resonance is associated with the attractive-exciton-polaron. In
contrast, the trion transition should have vanishing overlap with
the 2DES ground state and should not lead to strong coupling to the
cavity. We quantify the relative transition strength of the
attractive polaron and trion transitions in the discussion section
where we detail the theoretical model we used. We also emphasize
that the PL spectrum, which we associate with trion emission, shows
no normal mode splitting for the parameters which yield split
attractive-polaron-polariton peaks in transmission (Fig.~3f).

Figure~3d-f show cross sections through Figure~3a-c. The
simultaneous appearance of polaritons in both the repulsive and
attractive polaron branches (Fig.~3e) indicate that our experiments
make it possible to study a new regime of polaron physics where an
ultra-light polariton impurity is dressed with electron-hole pair
excitations. An exciting future direction motivated by our
observations is the investigation of polaron formation on polariton
transport.

The emergence of the lower-energy resonance and the gradual
disappearance of the exciton resonance as $n_e$ is increased has
been previously predicted and identified as oscillator strength
transfer from the exciton to trion
\cite{esser2001theory,smolka2014cavity}. To investigate how strong
coupling alters the polaron formation and the associated oscillator
strength transfer, we measured the normal mode splitting over a
large range of $V_g$. Figure~4a shows, as a function of $V_g$, the
transmission spectrum of the MoSe$_2$ monolayer when the bare cavity
mode is in resonance with the repulsive-polaron (left) or the
attractive-polaron. Increasing $n_e$ results in decreasing
(increasing) normal  mode splitting for the repulsive (attractive)
polaron. However, the maximum splitting for the attractive-polaron
obtained for $V_g \approx -5$~V is less than half as big as that of
exciton obtained in the absence of a 2DEG. Further increase in
$V_g$, or equivalently increase of $n_e$, results in diminishing
normal mode splitting. In the latter limit, the optical oscillator
strength is distributed over a broad energy range of order $E_F$,
thereby suppressing the coupling to the narrow cavity mode.

\paragraph*{}
For theoretical modeling we use  the truncated basis method (see
Supplementary Information), in which the Hilbert space is restricted
to include at most a single electron-hole pair \cite{Suris2003}.
Although an impurity inside a Fermi sea scatters an infinite number
of electron-hole pairs \cite{anderson1967infrared}, this variational
approach has been proven to be surprisingly accurate for modeling
cold-atom systems, due to the destructive interference of higher
order processes
\cite{vlietinck2013quasiparticle,combescot2008normal}. Since the
screened interaction between electrons and excitons is short ranged,
we treat it as a contact interaction. In contrast to the cold-atom
systems, the TMD-exciton-electron system has two types of excitons
and Fermi seas distinguished by the valley pseudospin
degree-of-freedom~\cite{xu2013TMDreview}. However, because at small
Fermi-energies, the interaction between excitons and electrons
inside the same valley is suppressed due to Pauli exclusion, we
neglect it altogether and model our system by a single excitonic
impurity in K (-K) valley interacting with electrons in the -K (K)
valley. We remark that electron-electron interactions can be
neglected if we truncate the Hilbert space to just one electron-hole
pair.  We also take into account phase-space filling effects, which
decrease the binding energy of the exciton and result in a blueshift
of all quasi-particle energies by $2 E_F$. This simple theoretical
model captures some but not all of the experimental observations in
MoSe$_2$. Finally, we note that for WSe$_2$ and WS$_2$, phase space
filling should be absent and as a consequence the attractive
(repulsive) polaron will exhibit a red (blue) shift
\cite{heinzPRL2015}.

We present the theoretical results  in the weak-coupling regime in
Figure~4b. As a function of $E_F$ all resonances exhibit a blueshift 
in energy that stems from phase-space filling. The exciton
resonance, which dominates the spectrum for vanishing $E_F$,
exhibits a further blueshift with increasing $E_F$, justifying its
identification as the repulsive polaron. For $E_F > 0$, a
lower-energy attractive polaron branch, with an energy comparable to
the molecule/trion energy, emerges~\cite{parish2011polaron}. In
contrast to the experimental results, our simple theoretical model
shows an abrupt turn-on of the blueshift of the repulsive polaron
branch for small $E_F$: this discrepancy possibly stems from the
disorder in the flake that results in localized electronic states
below the conduction band edge. We also capture the quasi-particle
weight transfer from the repulsive polaron to the attractive polaron
and the broadening of the repulsive polaron. Our model also predicts
a trion-hole continuum of width $2/3 E_F$, of very small weight.
Figure~4c shows that the quasi-particle weight of the trion-hole
continuum remains smaller than that of the attractive polaron branch
by a  factor of 10, supporting our claim that the polariton
formation cannot be associated with trions due to the small overlap
between the latter and the 2DES ground state + one cavity-photon. In
contrast to the negatively-charged trion, a Fermi polaron described
by the Chevy ansatz corresponds to a neutral excitation, consisting
of a Fermi-sea electron-hole pair bound to an
exciton~\cite{Suris2003}.

In our model, the $k=0$ trion is always higher in energy than the
attractive polaron. An extended truncated basis approach, which
includes states containing a molecule accompanied by an
electron-hole pair, can be used to show that the $k=0$ trion state
should be lower in energy for $E_F<5$~meV \cite{parish2011polaron}.
In contrast to the experimental findings, we cannot capture
correctly the broadening of the attractive polaron, since we have
not considered an ansatz with an extra electron-hole pair. Figure~4d
shows the calculated spectral function in the strong-coupling regime. We see that the polaron peaks become sharper due to the coupling to the narrow cavity. We notice that  we capture the
decrease (increase) in the light matter coupling for the attractive
(repulsive) polarons as $E_F$ increases. However, our model does not
predict the   full disappearance of the repulsive (or attractive)
polaron strong coupling to the cavity. Theoretically, at $E_F=30$~meV
the normal mode splitting of the repulsive polaron is reduced to 7~meV while the normal mode splitting of the attractive polaron is
roughly 13~meV. The discrepancies between the experimental data
and the theoretical predictions may also stem from our approximation
of a rigid exciton: our model does not capture exchange and
correlation effects which are known to play a role in trion
formation~\cite{esser2001theory}.

Our experiments establish strongly bound excitons in TMD monolayers,
simultaneously embedded in a 2DES and a microcavity, as a new
paradigm for quantum impurity and polaron physics. In stark contrast
to prior work, we identify the optical excitations that are
accessible in resonant spectroscopy as repulsive and attractive
exciton polarons and polaron-polaritons, which are simultaneously
present for Fermi energies that are smaller than the molecular
(trion) binding energy. For $E_F$ exceeding the conduction band
spin-orbit coupling, TMD monolayers exhibit both intra- and
inter-valley trions that are coupled by electron-hole
exchange~\cite{wangyao2014}: an interesting open question is whether
the Berry curvature of the corresponding exchange coupled intra- and
inter-valley attractive polarons leads to novel transport
signatures. While we report the measurement of the spectral function
of the interacting polariton-2DES system, we highlight that it is
possible to directly measure the nonequilibrium response of the
system in the time domain using ultrashort laser pump-probe
spectroscopy in the regime $E_F \le 10$~meV. Finally, another
interesting extension of our work would be the investigation of a
Bose-polaron problem where an optically injected -K valley polariton
impurity interacts with Bogoliubov excitations out of a polariton
condensate in the +K valley.

\paragraph*{\textbf{Acknowledgements}}
J. Reichel, A. Kis and R. Schmidt have made invaluable contributions
to the experimental and theoretical aspects of this work.  The authors also
acknowledge many insightful discussions with C. Ciuti, M. Combescot,
M. Fleischhauer, L. Glazman, M. Goldstein, F. Grusdt, D. Pimenov and
J. von Delft. This work is supported by an ERC Advanced investigator
grant (POLTDES), Harvard-MIT CUA, NSF Grant No. DMR-1308435, Dr.~Max R\"ossler, the Walter Haefner Foundation and the ETH Foundation.

\clearpage

\onecolumngrid

\begin{figure}
	\begin{center}
		\includegraphics[scale=1.0]{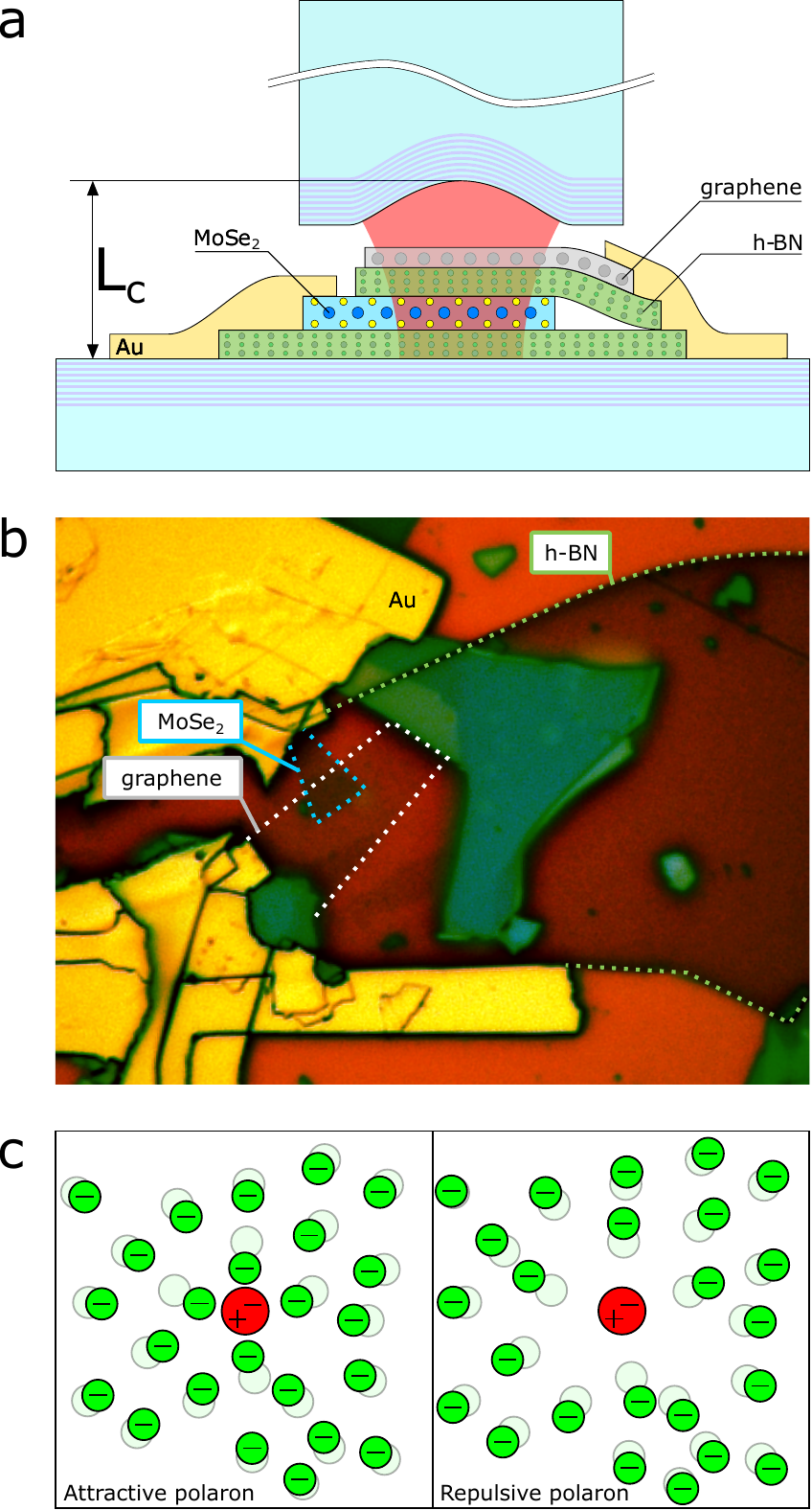}
		\caption{\textbf{A MoSe$_2$/hBN/graphene heterostructure in a fiber
			cavity.} \textbf{a}, The sample consists of a $3\, \mu$m by $5\, \mu$m
			MoSe$_2$ monolayer sandwiched between $10$nm and $110$nm thick hBN
			layers. A graphene layer on top completes the heterostructure that
			allows for controlling the electron density in the MoSe$_2$
			monolayer by gating. The heterostructure is placed on top of a flat
			dielectric mirror (DBR). The thicknesses of the hBN layers are
			chosen to ensure that the MoSe$_2$ monolayer is at an antinode and
			the graphene layer is at a node of the cavity formed by the bottom
			dielectric mirror and the top fiber mirror. The finesse of the
			cavity is $\sim 200$; the cavity length can be tuned from $1.9\, \mu$m
			to $15 \,\mu$m. \textbf{b}, The optical microscope image of the
			heterostructure where the overlap between the MoSe$_2$ monolayer and
			the top graphene  layer is identified. \textbf{c}, Due to
			exciton-electron interactions, the exciton is surrounded by an
			electron screening cloud that leads to the formation of an
			attractive polaron (left panel). For a repulsive polaron the
			electrons are pushed away from the exciton leading to a higher
			energy metastable excitation (right panel).}
	\end{center}
\end{figure}

\begin{figure}
	\begin{center}
		\includegraphics[scale=0.3]{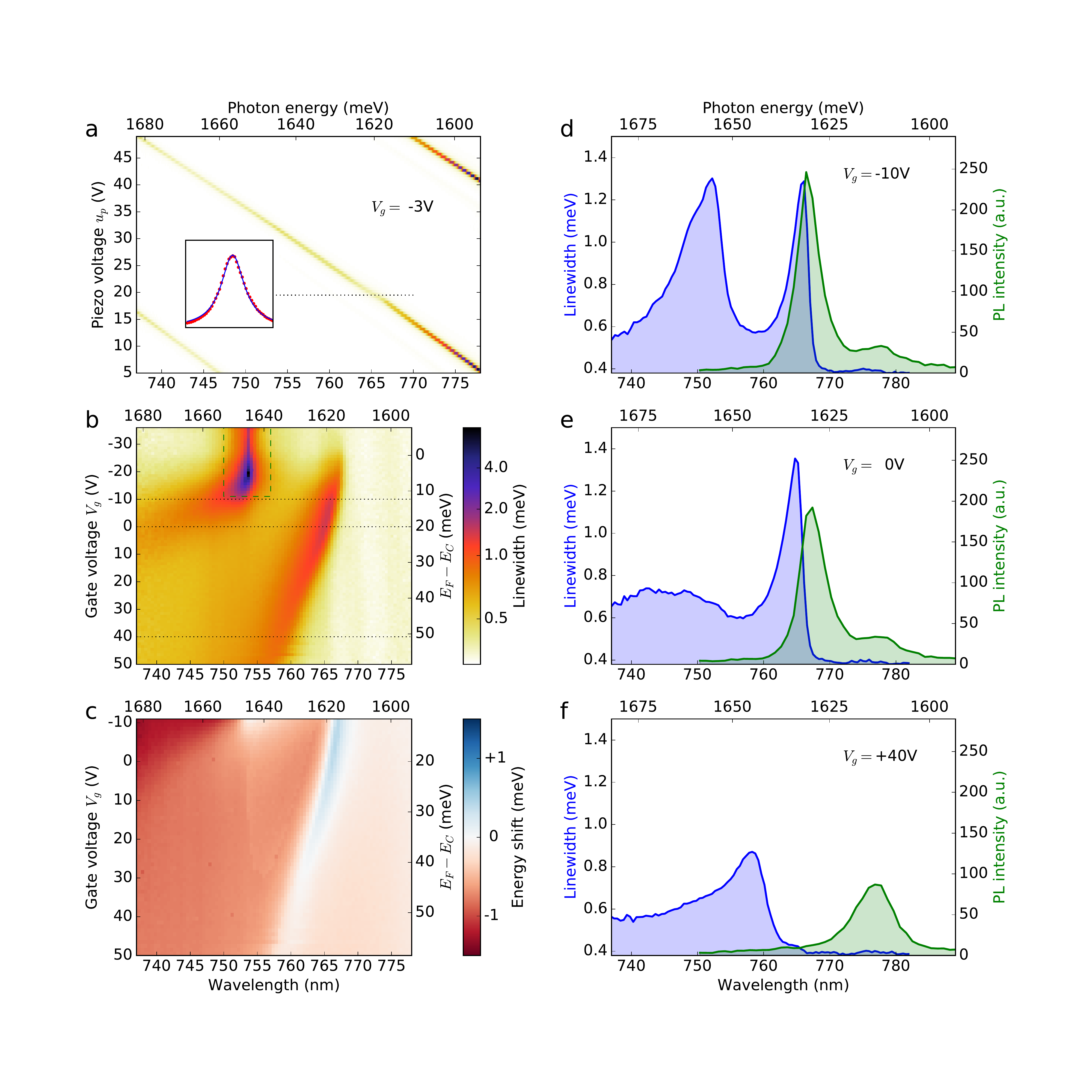}
		\caption{\textbf{Cavity spectroscopy of the interacting
			exciton-electron system in the weak coupling regime.} \textbf{a},
			The white light transmission spectrum of the fiber cavity
			incorporating  the MoSe$_2$/hBN/graphene heterostructure, as a
			function of the piezo voltage (vertical scale) that is varied to
			tune the cavity frequency. Since the bare cavity linewidth of
			$0.3$meV is much smaller than all other energy scales, cavity
			transmission allows for identifying the linear optical response of
			the heterostructure: Whenever the cavity mode is at a frequency
			absorbed by the MoSe$_2$ flake, its linewidth increases.
			Consequently, the MoSe$_2$ absorption spectrum can be measured as a
			frequency dependent broadening of the cavity. The insert shows the
			cavity transmission at $V_g = -3$~V and $u_p = 20$~V fitted with a
			lorentzian curve. \textbf{b}, The MoSe$_2$ absorption spectrum
			determined by measuring the enhancement of the cavity linewidth for
			each cavity frequency (horizontal axis) and gate voltage or
			equivalently the Fermi energy (vertical axis). When the Fermi energy
			is below the conduction band minimum ($E_C$, we choose $E_C = 0$), absorption is only observed at
			the bare exciton frequency. As electrons are introduced, the exciton
			resonance experiences a sharp blueshift together with broadening.
			Concurrently, there is a new resonance emerging at $\sim 25$~meV
			below the bare exciton energy. \textbf{These features are identified
				as the repulsive and attractive exciton-polaron resonances.} For
			$V_g < -10$V, the exciton and the cavity mode are in the
			strong-coupling regime (the region highlighted using the dashed
			rectangle) and it is not possible to directly extract imaginary part
			of the MoSe$_2$ linear susceptibility: in this regime, we measure
			and plot the linewidth of the cavity-like polariton. \textbf{c}, The
			measured real part of the susceptibility of the MoSe$_2$ flake as  a
			function of the cavity frequency (horizontal axis) and the gate
			voltage (vertical axis). The data presented here is connected to the
			absorption data of Fig.~2b via Kramers-Kroenig relations. In the
			absence of the MoSe$_2$ flake, there is an expected change of the
			cavity resonance (peak in the spectrum of the transmitted light)
			with changing piezo voltage. In the presence of MoSe$_2$, the index
			of refraction seen by the photons is modified due to the real part
			of the MoSe$_2$ susceptibility, thereby modifying the effective
			cavity length and leading to a shift of the cavity resonance
			wavelength as compared to what we would have obtained in the absence
			of MoSe$_2$. \textbf{d}, Line-cut through the cavity line broadening
			data (blue shaded curve) for $V_g = -10$~V: both repulsive and
			attractive polaron features are visible. We expect the trion and
			attractive-polaron energies to be comparable for this $V_g$. The
			photoluminescence (PL) data is shown in green. \textbf{e}, Line-cut
			through the cavity line broadening and shift data for $V_g = 0$~V:
			the line broadening/absorption data is dominated by the attractive
			polaron which is now blue-shifted with respect to the trion PL.
			\textbf{f}, Line-cut through the cavity line broadening and shift
			data for $V_g = 40$~V: the PL and absorption peaks are separated
			from each other by $40$~meV suggesting that PL and absorption data
			stem from different quasiparticles, namely the trion and the
			attractive polaron, respectively.}
	\end{center}
\end{figure}

\begin{figure}
	\begin{center}
		\includegraphics[scale=0.3]{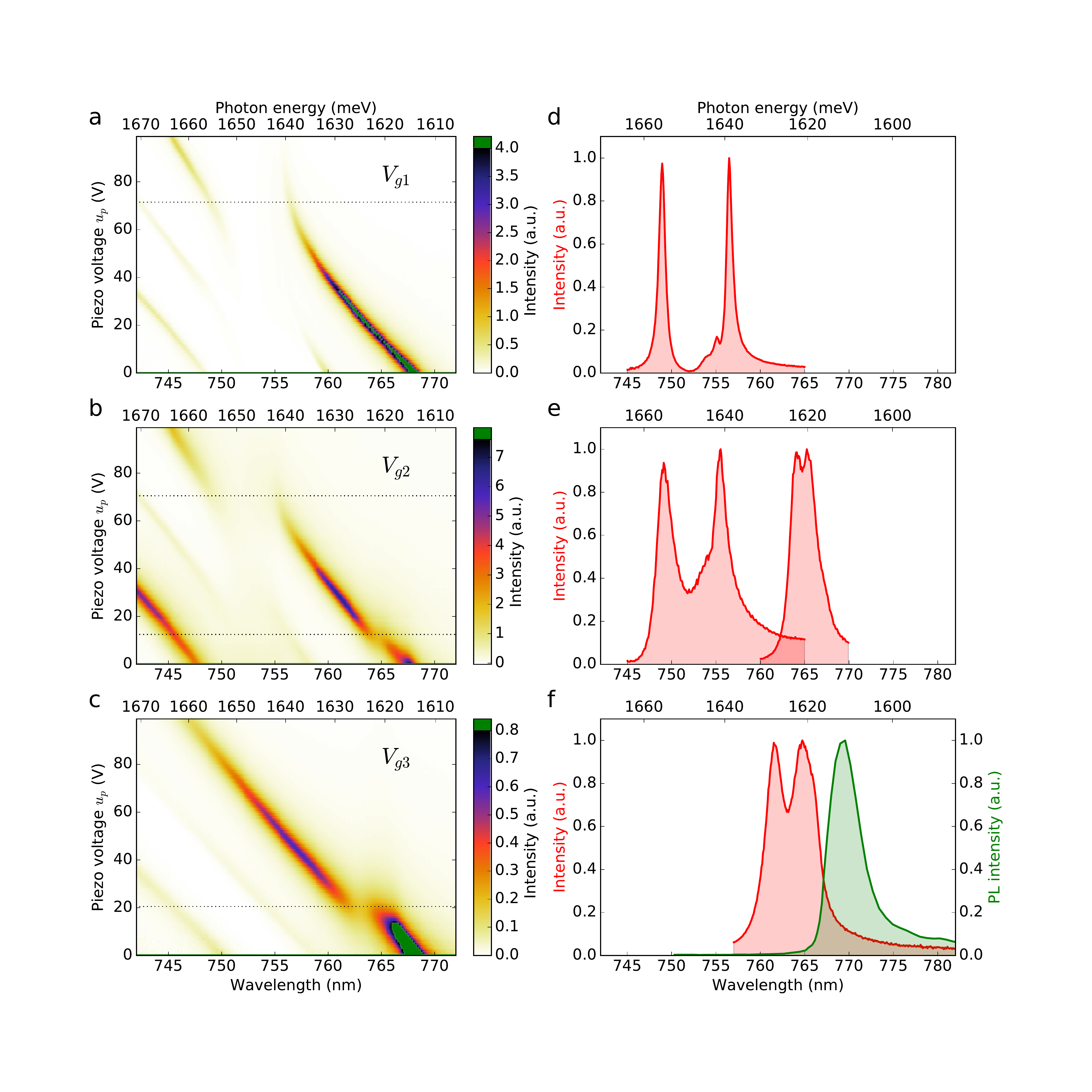}
		\caption{\textbf{Cavity spectroscopy of the interacting
			exciton-electron system in the strong coupling regime.} \textbf{a},
			The white light transmission spectrum as a function of the piezo
			voltage (vertical scale) for an average cavity  length of $1.9
			\,\mu$m. Due to enhanced cavity electric field, the interaction
			between the cavity mode and MoSe$_2$ resonances is directly observed
			in cavity transmission spectra as anticrossings associated with
			polariton formation. For gate voltages where the MoSe$_2$ monolayer
			is devoid of electrons ($V_{g1}$) the spectrum shows a prominent
			anticrossing with a normal mode splitting of $16$~meV. The
			elementary optical excitations in this regime are bare
			exciton-polaritons without any polaron effect. The green area indicates values outside the range of the colormap. \textbf{b}, White
			light transmission spectrum for $V_g = V_{g2} = -5$~V, showing two
			anticrossings associated with the formation of repulsive- and
			attractive-polaron-polaritons. \textbf{The observation of
			anticrossings for both lower and higher energy resonances proves
			that these originate from Fermi-polarons with a large quasiparticle
			weight}. \textbf{c}, White light spectrum for a higher gate voltage
			($V_{g3}$), where only the attractive-polaron exhibits
			non-perturbative coupling to the cavity mode.  \textbf{d} Line-cut
			through the data in Fig.~3a for the piezo voltage $u_p = 72$~V shows
			the transmission spectrum (red curve) at the resonance of the cavity
			with the exciton. \textbf{e}, Line-cut through the data in Fig.~3b
			for the piezo voltage $u_p = 13$~V respectively $u_p = 71$~V.
			\textbf{f} Line-cut through the data in Fig.~3e for $u_p = 21$~V,
			corresponding to the case where the cavity mode is resonant with the
			attractive polaron resonance. The photoluminescence spectrum in the
			strong coupling regime is also plotted (green shaded curve).}
	\end{center}
\end{figure}

\begin{figure}
	\begin{center}
		\includegraphics[scale=0.3]{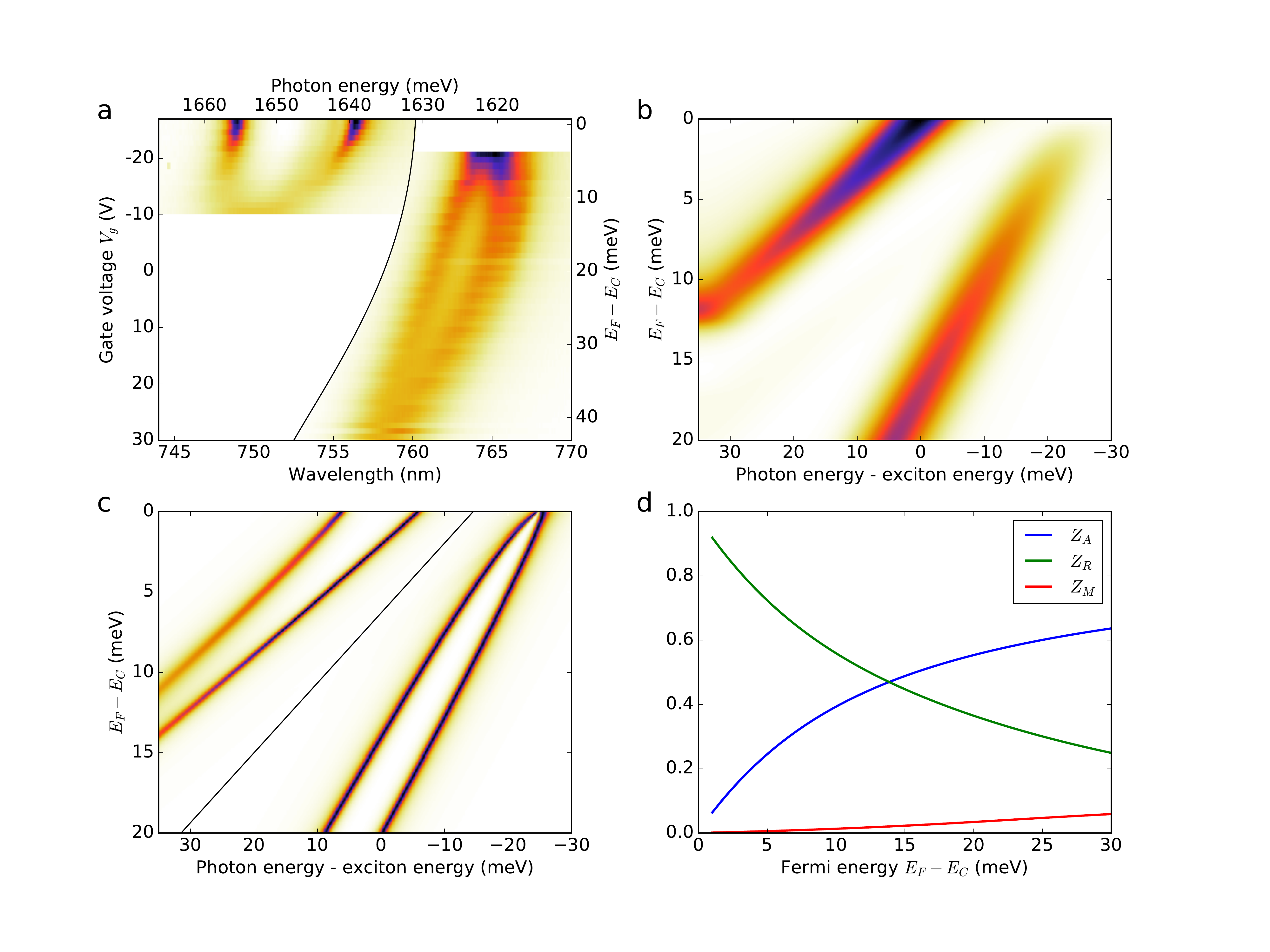}
		\caption{\textbf{Competition between repulsive and attractive
			polaron resonances.} \textbf{a}, The white light transmission
			spectrum  of the fiber cavity incorporating the
			MoSe$_2$/hBN/graphene heterostructure, as a function of the gate
			voltage (vertical scale) for two different settings of the cavity
			length: the left (right) part shows the transmission when the cavity
			is tuned on resonance with the repulsive (attractive) polaron. For
			each horizontal line, the cavity frequency is tuned so as to yield
			two polariton modes with equal peak amplitude. As $V_g$ is
			increased, the oscillator strength transfer from the repulsive to
			attractive branch is clearly visible. While the normal mode
			splitting for the repulsive branch disappears for $V_g \simeq -10$~V,
			the collapse of the splitting takes place at $V_g = 25$~V for the
			attractive branch. \textbf{b}, The spectral function calculated
			using the Chevy ansatz in the weak coupling regime is in good
			qualitative agreement with the absorption spectra of Fig.~2b.
			\textbf{c}, The calculated quasi-particle weights showing the
			oscillator strength transfer from the repulsive to the attractive
			polaron as the Fermi energy is increased. The weight of the
			trion+hole continuum increases linearly with the Fermi energy but
			remains less than 0.1 even for Fermi energies exceeding the trion
			binding energy. \textbf{d}, The spectral function calculated using
			the Chevy ansatz in the strong coupling regime captures the
			oscillator strength transfer from the repulsive to attractive
			polaron depicted in Fig.~4a but fails to predict the collapse of the
			normal mode splitting with increasing electron density.}
	\end{center}
\end{figure}

\clearpage

\section{ Supplementary Materials: Fermi polaron-polaritons in charge-tunable atomically thin semiconductors}

\setcounter{equation}{0}
\setcounter{figure}{0}
\setcounter{table}{0}
\setcounter{page}{1}
\makeatletter
\renewcommand{\theequation}{S\arabic{equation}}
\renewcommand{\thefigure}{S\arabic{figure}}
\renewcommand{\bibnumfmt}[1]{[S#1]}
\renewcommand{\citenumfont}[1]{S#1}

\section{Theory}
Although in the experiments we have a 0D cavity, these experiments can be extended to a system with a 2D cavity. Therefore, in the theory section we will analyse the latter case. The 0D cavity case will appear as a special case of our result. However, in considering a 2D cavity we show that ultra-low mass polarons can easily be obtained experimentally by exchanging the 0D cavity with the 2D cavity.

We start from the following Hamiltonian:
\bea
\label{seq1}
H &=&\sum_k \omega_C(k) c^{\dagger}_k c_k + \sum_k \omega_X(k) x_k^{\dagger} x_k +  \sum_k g (c_k^{\dagger} x_k + h.c.) +  \sum_k \epsilon(k) e_k^{\dagger} e_k + \sum_{k,k',q} V_q x_{k+q}^{\dagger} e_{k'-q}^{\dagger} e_{k'} x_{k} \\
\omega_C(k) &=& \frac{\hbar k^2}{2 m_c}, \quad \omega_X(k) = \frac{\hbar k^2}{2 m_{exc}}+2 E_F, \quad \epsilon^{(e)}_k = \frac{\hbar k^2}{2 m_e},
\eea
where $c,x$ and $e$ are the destruction operators of a cavity photon, an exciton and an electron respectively, while $m_c,m_{exc}$ and $m_e$ are the masses of the cavity photon, the exciton and the electron. The third term corresponds to the coupling between excitons and the cavity field, while the last term incorporates the interaction between the exciton and the Fermi sea. Although we consider the exciton to be a rigid object, the interaction between the exciton and the Fermi sea contains the effects due to electron exchange between the exciton and the Fermi sea. The $2E_F$ term in the exciton dispersion is due to phase space filling which results in an overall blueshift of the exciton line.

\subsection{Chevy  ansatz for the polaron}
In order to analyze the problem we make a Chevy-type ansatz
\cite{chevy2006universal} for the polaron state, which truncates the
Hilbert space to a single electron-hole pair: \bea | \Psi  ^{(p)}
\rangle = \left(\phi_0 x_p+\varphi_0  c_p + \sum_{k,q} \phi_{k,q}
x^\dagger_{p+q-k} e^\dagger_{k} e_q +\sum_{k,q} \varphi_{k,q}
c^\dagger_{p+q-k} e^\dagger_{k} e_q\right) | 0 \rangle, \eea where we
defined the vacuum $|0\rangle$ as an undisturbed Fermi sea and no
excitons  in the system. This ansatz takes into account the total
momentum conservation in our system and describes a quasi particle
of momentum $p$ formed by the superposition of a cavity photon and
an exciton dressed by an electron-hole pair from the Fermi sea. To
obtain the ground-state we must minimize the quantity  $ \langle
\Psi^{(p)} | E - H | \Psi^{(p)} \rangle $: \bea
\langle \Psi^{(p)} | E - H | \Psi^{(p)} \rangle &=& E \left(|\phi_0|^2 +  |\varphi_0|^2 +\sum_{k,q} | \phi_{k,q}|^2 +\sum_{k,q} | \varphi_{k,q}|^2 \right) - H_{\mathrm{var}}^{(p)}
\eea

\begin{align}
\begin{split}
H_{\mathrm{var}}^{(p)}&=\langle \Psi^{(p)}| H | \Psi^{(p)}\rangle =\omega_X(p)  |\phi_0|^2  + \omega_C(p) |\varphi_0|^2 + \sum_{k,q} E_X(p,k,q)| \phi_{k,q}|^2+ \sum_{k,q} E_C(p,k,q) |\varphi_{k,q}|^2 \\
& - g \left[\phi_0^* \varphi_0+\sum_{k,q}\phi_{k,q}^* \varphi_{k,q}+c.c.\right] + |\phi_0|^2 \sum_q V_0 + \sum_{k,q} \left[   \phi_0  ^* \phi_{k,q}V_{k-q} + c.c. \right] \\
&+\sum_{k,q,k'} \left[ \phi_{k,q}^* \phi_{k',q} V_{k-k'}+c.c.\right]-\sum_{k,q,q'} \left[ \phi_{k,q} ^* \phi_{k,q'}V_{q-q'}+c.c.\right],
\end{split}
\end{align}

where $E_X(p,k,q)\equiv\omega_X(p+q-k)+\epsilon(k)-\epsilon(q)$ and  $E_C(p,k,q)\equiv\omega_C(p+q-k)+\epsilon(k)-\epsilon(q)$.
Each term in $H_{var}$ corresponds to a physical process allowing the observation of the  competition between different processes, in trying to minimize $H_{var}$ subject to the normalization constraint. Minimizing the above equation, we obtain the following equations:
\bea
& &\omega_C(p)\varphi_0  -g \phi_0= E \varphi_0\\
& &E_C(p,k,q)\varphi_{k,q}  - g \phi_{k,q} = E \phi_{k,q} \\
& & \omega_X(p)\phi_0  -g \varphi_0 +\sum_{q} \phi_0  V_0 +\sum_{k,q} \phi_{k,q} V_{k-q} = E \phi_0\\
& & E_X(p,k,q) \phi_{k,q} - g\varphi_{k,q}  + V_{k-q} \phi_0 + \sum_{k'} V_{k'-k}\phi_{k',q}+ \sum_{q'} V_{q'-q}\phi_{k,q'} = E \phi_{k,q}.
\eea
Because the cavity coupling does not mix different momentum states, we can eliminate the first two equations and obtain a set of two equations:
\bea
& & \left(\omega_X(p) -\frac{g^2}{\omega_C(p)} \right) \phi_0  +\sum_{q} \phi_0  V_0 +\sum_{k,q} \phi_{k,q} V_{k-q} = E \phi_0\\
& &\left(E_X(p,k,q) - \frac{g^2}{E_C(p,k,q)}\right) \phi_{k,q}  + V_{k-q} \phi_0 + \sum_{k'} V_{k'-k}\phi_{k',q}+ \sum_{q'} V_{q'-q}\phi_{k,q'} = E \phi_{k,q}   \label{seq3}.
\eea
Notice that the effect of the cavity is to renormalize the energies of many body states. The correction is recognized as the exact self-energy due to interactions with the cavity field.

At this point it is straightforward to solve the problem by discretizing the momenta $k,q$ and transforming the above equations into a matrix equation.

However, we can make further analytical progress by making a few reasonable approximations. First of all, we notice that the exciton-electron interaction is a Van-der-Waals interaction which decays as $r^{-4}$ at large distances, with a range given by the Bohr radius $a_B$ of the excition. Including the screening effects due to the electron system, the interaction will become even shorter range. Therefore, at least for small Fermi energies (i.e. $a_B k_F \ll 1$) we can approximate the interaction with a contact interaction which is constant $V_k=V$ up to a cutoff $\Omega$. Since in two dimensions, an attractive potential always has a bound state of energy, in our case we denote it by $-E_T$, we can express the interaction strength as a function of the bound state energy and an ultraviolet cutoff:
\bea
\frac{1}{V}=-\sum_{k=0}^{\Omega}\frac{1}{E_T-\omega_X(0)+\omega_X(k)+\epsilon(k)}.
\eea
Since the physics should not depend on the ultraviolet cutoff,  in the end we will let $\Omega \to \infty$ and therefore $V \to 0$. As we will show a posteriori, $\phi_{k,q} \sim 1/k^2$ for large $k$, which in turn implies that the last term on the left hand side vanishes when $V \to 0$. We will therefore proceed by neglecting this term.

We introduce the function $\chi_q = \phi_0 + \sum_k \phi_{k,q}$. In terms of this function:
\bea
\phi_0 &=& \frac{V \sum_q \chi_q}{E-\omega_X(p)+\frac{g^2}{\omega_C(p)}}\\
\phi_{k,q}&=& \frac{V \chi_q}{E-E_X(p,k,q) + \frac{g^2}{E_C(p,k,q)}}.
\eea
Reintroducing the above into the definition of $\chi_q$ we can obtain the following self-consistent equation:
\bea
E =\omega_X(p) + \frac{g^2}{E-\omega_C(p)} + \sum_q \left[ \sum_{k=0}^{\Omega} \frac{1}{E_T-\omega_X(0)+\omega_X(k)+\epsilon(k)} -\sum_{k=k_F}^{\Omega}\frac{1}{E-E_X(p,k,q)+\frac{g^2}{E-E_C(p,k,q)}} \right]^{-1}.
\eea
To gain further insight into the above equation we introduce the dispersion resulting from linearly coupling two harmonic oscillator modes of energies $E_X(p,k,q)$ and $E_C(p,k,q)$ with a coupling strength $g_c$. These resemble the polariton modes:
\bea
\Omega_{LP,UP}(p,k,q)&=&\frac{1}{2}\left[ E_X(p,k,q)+E_C(p,k,q) \pm \sqrt{\left(E_X(p,k,q)-E_C(p,k,q)\right)^2+4 g_c^2}\right].
\eea
We also introduce the factors resembling the exciton fractions in polaritons. which show how much of the initial modes is contained in the new modes:
\bea
| X(p,k,q) |^2&=& \frac{1}{2} \left(1 + \frac{E_C(p,k,q)-E_X(p,k,q)}{\sqrt{\left(E_C(p,k,q)-E_X(p,k,q)\right)^2 + 4 g_c^2}} \right).
\eea
With the above notation we can rewrite the self consistent equation as:
\begin{align}
\begin{split}
E - \omega_X(p) = & \frac{g^2}{E-\omega_C(p)}  \\ +  &\sum_q \left[ \sum_{k=0}^{\Omega} \frac{1}{E_T-\omega_X(0)+\omega_X(k)+\epsilon(k)} -\sum_{k=k_F}^{\Omega} \left(\frac{|X(p,k,q)|^2}{E-\Omega_{LP}(p,k,q)} + \frac{1-|X(p,k,q)|^2}{E-\Omega_{UP}(p,k,q)} \right) \right]^{-1}.
\end{split}
\end{align}
We can simplify things further by noting that for $p+q-k>k_{ph}$ ($k_{ph}$ is of the order of the photon momentum and approximately given by $\hbar k_{ph}^2/(2 m_c)=g_c$) $X(p,k,q) \approx 1$ and $\Omega_{LP}(p,k,q) \approx E_X(p,k,q)$. Since $k_{ph}$ is much smaller than all the other momentum scales the phase space where these approximations break down is extremely small. Based on this phase-space argument we can simplify the above equation:
\bea
E  =\omega_X(p)+ \frac{g^2}{E-\omega_C(p)} + \sum_q \left[ \sum_{k=0}^{\Omega} \frac{1}{E_T-\omega_X(0)+\omega_X(k)+\epsilon(k)} -\sum_{k=k_F}^{\Omega} \frac{1}{E-E_X(p,k,q)}   \right]^{-1}
\eea
We would have obtained the same equation if we started from an ansatz which did not contain the states corresponding to a photon dressed by an electron-hole pair (i.e. $\varphi_{k,q}=0$). Our full derivation serves to justify this approximation. We remark that the poles in the $q$ summation correspond to the molecular energies that are obtained when choosing an ansatz of the form $|\Phi^{(p)}\rangle =  \phi_{k} x^\dagger _{p-k} e^\dagger_k e_p |0\rangle$.

It can be shown that by replacing $E \to E + i \eta $ ($\eta \to 0^+$), the last term on the right hand side of the above equation is the self-energy of an exciton interacting with a Fermi sea. Although the inclusion of the infinitesimal $i\eta$ might seem arbitrary at this point, it can be shown that it emerges from choosing a time dependent ansatz, and instead of minimizing $\langle \Phi | H | \Phi \rangle $, minimizing the action $S=\int \langle  \Phi (t) | i\partial/\partial t - H | \Phi(t) \rangle$ \cite{Sparish2013highly}. Therefore, the above equation can be written more intuitively as:
\bea
E &=& \omega_X(p) + \Sigma_X(E,p) \\
\Sigma_X(E,p) &=& \Sigma_{X-C}(E,p)  +\Sigma_{X-e}(E,p) \\
\Sigma_{X-C}(E,p)&=&\frac{g^2}{E-\omega_C(p)}\\
\Sigma_{X-e}(E,p)&=&\sum_q \left[ \sum_{k=0}^{\Omega}
\frac{1}{E_T-\omega_X(0)+\omega_X(k)+\epsilon(k)} -\sum_{k=k_F}^{\Omega}
\frac{1}{E+i \eta -E_X(p,k,q)}   \right]^{-1} \eea
In the above we
made explicit the self energy of the exciton interacting with the
cavity mode ($\Sigma_{X-C}$) and with  the electrons in the Fermi
sea ($\Sigma_{X-e}$).

Having found the self-energy of the exciton, we can also obtain the
self-energy of the cavity photon: \bea \Sigma_C (E,p) =
\frac{g^2}{E - \omega_X(p)-\Sigma_{X-e}(E,p)} \eea .

\subsection{Spectral Function}
In the weak-coupling regime, in our experiment, we are probing the exciton spectral function:
\bea
A(t) = \langle 0 | x_0 e^{-i H t} x_0^\dagger |0\rangle
\eea
In the truncated basis the Fourier transform of the spectral function is given by:
\bea\label{WK}
A(\omega) = \frac{1}{\pi} \mathrm{Im}\left[ \frac{1}{\omega+i\eta-\omega_0^{(x)}-\Sigma_X(\omega,0)} \right]
\eea

In the strong coupling regime we are probing the spectral function of the cavity photon and therefore, in the above we should replace $\Sigma_X$ with $\Sigma_C$.

In simulating the experimental results we choose a lifetime
broadening of the exciton/photon linewidth of $\eta = 1.0$~meV.
Since the exciton is also subject to disorder broadening, in the
weak coupling regime we convolve the resulting spectral function
with a Gaussian kernel with a standard deviation of 14~meV (such
that FWHM~=~7~meV), obtained from fitting the experimental exciton line at zero
Fermi energy.

\subsection{Polaron mass}
Having found the exciton self-energy we can also determine the effective mass of the exciton due to the interaction with the light cavity photon and the electron system. Assuming that the lowest energy state is at energy $E_0$ and momentum 0, the effective mass is given by:
\bea
\frac{1}{m^*} = \frac{1}{m_{x}}+\frac{\partial}{\partial^2 p} \Sigma_X(E_0,p) \Bigg|_{0}= \frac{1}{m_{x}}+\frac{1}{m_{c}}+\frac{\partial}{\partial^2 p} \Sigma_{X-e}(E_0,p) \Bigg|_{0}
\eea
Regardless of the contribution of the last term in the above equation we can see that, due to the small mass of the cavity photon, the polaron mass is going to be ultra-small. Therefore, we conclude that we are dealing with an ultra-low mass polaron. We emphasize that an ultra-low mass polaron can only be achieved by dressing a (polariton) impurity which is a superposition of an ultra-low mass particle (cavity photon) with a relatively heavy particle (exciton). Such a mixed-impurity exhibits an ultra-small mass for low momenta but restricts the recoil energy to the coupling energy $g_c$. Otherwise, if we did not have the relatively heavy particle, the dressing of an ultra-low mass impurity is very ineffective since $E_T \to 0$ for the same $V$. This means that the impurity will not be affected by the Fermi sea at all.

\section{Sample preparation and measurement setup}

The heterostructure studied in this work was  assembled using the
pickup technique \cite{zomer2014}: Flakes of the constituent materials are exfoliated
onto separate substrates and sequentially picked up with a polycarbonate layer which finally deposits the complete heterostructure
onto the target substrate. The target substrate in this case is a
distributed Bragg reflector (DBR) ion beam sputtered onto a fused silica substrate. The DBR is
designed to have a reflectivity of $> 99.3\%$ for the spectral range
of 680-800 nm and an intensity maximum at the DBR surface. The
graphene top gate as well as the MoSe$_2$ flake were contacted using
metal gates consisting of a thin layer of titanium followed by a
thicker layer of gold. In order to increase the chance of a good
contact to the MoSe$_2$ flake, a parallel contact via a second
graphene flake was made.

The top mirror is formed by a dimple with  radius of curvature of
$30\, \mu$m shot into the fiber facet with a CO$_2$ laser. The
geometry of the dimple was measured with interferometry. The fiber
facet was coated with the same DBR as described above. All
measurements are performed using a dipstick immersed in liquid
helium. The sample can be moved in the (x-y) plane using
nanopositioners. In addition, the cavity length is adjusted with a
z-axis nanopositioner. For transmission, light from a
(broadband) LED covering the spectral range of interest is sent
through the fiber with the dimple and collected by an aspheric lens.
A second LED emitting green light, that overlaps with a transmission
window of the DBR mirrors, is used to locate the flake.

For PL measurements, a 532~nm laser is sent through the fiber. This
wavelength is within a transmission window outside of the stop band
of the DBR. Therefore, PL excitation is efficient and only marginally changing
with the cavity length. To extract the PL spectrum, the cavity emission spectrum
is measured as a function of cavity length. For each cavity length, the area and the
center wavelength of the PL escaping through the cavity mode is measured.
Plotting the area against the center wavelength of the cavity mode yields the
PL spectrum of the flake.

\section{Capacitive model for the Fermi Energy}

By applying a top gate voltage $V_g$ the electron density in the sample and therefore the Fermi energy $E_F$ is changed. We denote the smallest $V_g$ for which the attractive polaron is observed as $V_g = V_c$ which we interpret as the gate voltage for which we start populating the conduction band ($E_F > 0$).

The capacitance per unit area $C/A$ between top gate and sample is given by:

\bea
\frac{C}{A} = \left(\frac{t}{\epsilon \epsilon_0} + \frac{1}{e^2\mathrm{D}(E)}\right)^{-1},
\eea

where $\mathrm{D}(E) $ is the density of states and $t$, $\epsilon$ are the thickness respectively the permittivity of the hBN flake. The two terms are the geometric respectively quantum capacitance of the sample. For $E_F > 0$, the quantum capacitance can be neglected since its effect is much smaller and within the uncertainty of the permittivity of the hBN flake. For $V_g > V_c$ i.e. $E_F > 0$ this yields:

\bea
E_F = \frac{\pi \hbar ^2 \epsilon \epsilon_0}{t e m^*}(V_g-V_c) \approx 0.77 \frac{\text{meV}}{\text{V}}(V_g - V_c),
\eea

where $m^*$ is the effective electron mass of the conduction band.

\section{Spatial dependence}

One of the principal advantages of the open fiber  cavity structure
is the ability to adjust the cavity length and to thereby change the
nature of coupling. In our setup, the fiber facet and the substrate
form a small angle. As a consequence, when the fiber is in close
proximity to the facet it touches the substrate. The
contact is located at the edge of the fiber which is $125\, \mu$m in
diameter. This contact stabilizes the cavity by suppressing the
vibrations that would otherwise have lead to line broadening.
Furthermore, once the cavity is brought into contact with the
substrate, changing the cavity length by changing the piezo voltage
of the z-axis nanopositioner seems to be completely reversible. When
the cavity length is reduced further to a few micrometers, it is
essentially the fiber angle that changes and reduces the cavity
length at the dimple which is in the center of the fiber facet.

Additionally, scanning the sample with respect to the fiber mirror
allows us to  investigate the spatial dependence of the MoSe$_2$
optical excitations. In order to investigate the latter, the cavity
length was enlarged to be sure to eliminate any contact between fiber and
substrate. At a cavity length of $\sim 30 \mu$m, the sample was moved
with respect to the fiber mirror with nanopositioner slip-stick
steps. The nanopositioner resistive readout was used to get an
estimate of the traveled distance. A rough estimate of the cavity
positon with respect to the flake is obtained using a camera by imaging the
flake illuminated with a green LED which is transmitted through a transmission
window of the DBR. The spectrum is derived from the cavity linewidth
broadening in a cavity length scan using the same technique as for
the data shown in Figure 2 of the main text.

Figure~S1 shows absorption spectra for different positions of the sample with respect to the cavity. The sample is moved by $\sim 0.5\, \mu$m in between the different measurements. The scan was measured at a gate voltage of $V_g = -10$~V where we expect to see absorption from both the repulsive and attractive polaron. At $x = -1.5\, \mu$m, the spatial overlap of the cavity mode with the MoSe$_2$ monolayer is small. As the sample is moved, the overlap and therefore the absorption increases until it reaches a maximum at  $x = 0.0\, \mu$m. Moving the sample further up to $x = 1.5\, \mu$m reduces the overlap again. The absorption strength of the attractive polaron as compared to the repulsive polaron does not change significantly depending on the position of the sample which indicates a relatively homogeneous electron density. The large distance over which the absorption decreases is in accordance with the large cavity mode waist of $~1.7\, \mu$m.

\begin{figure}
\begin{center}
		\includegraphics[width = 10cm]{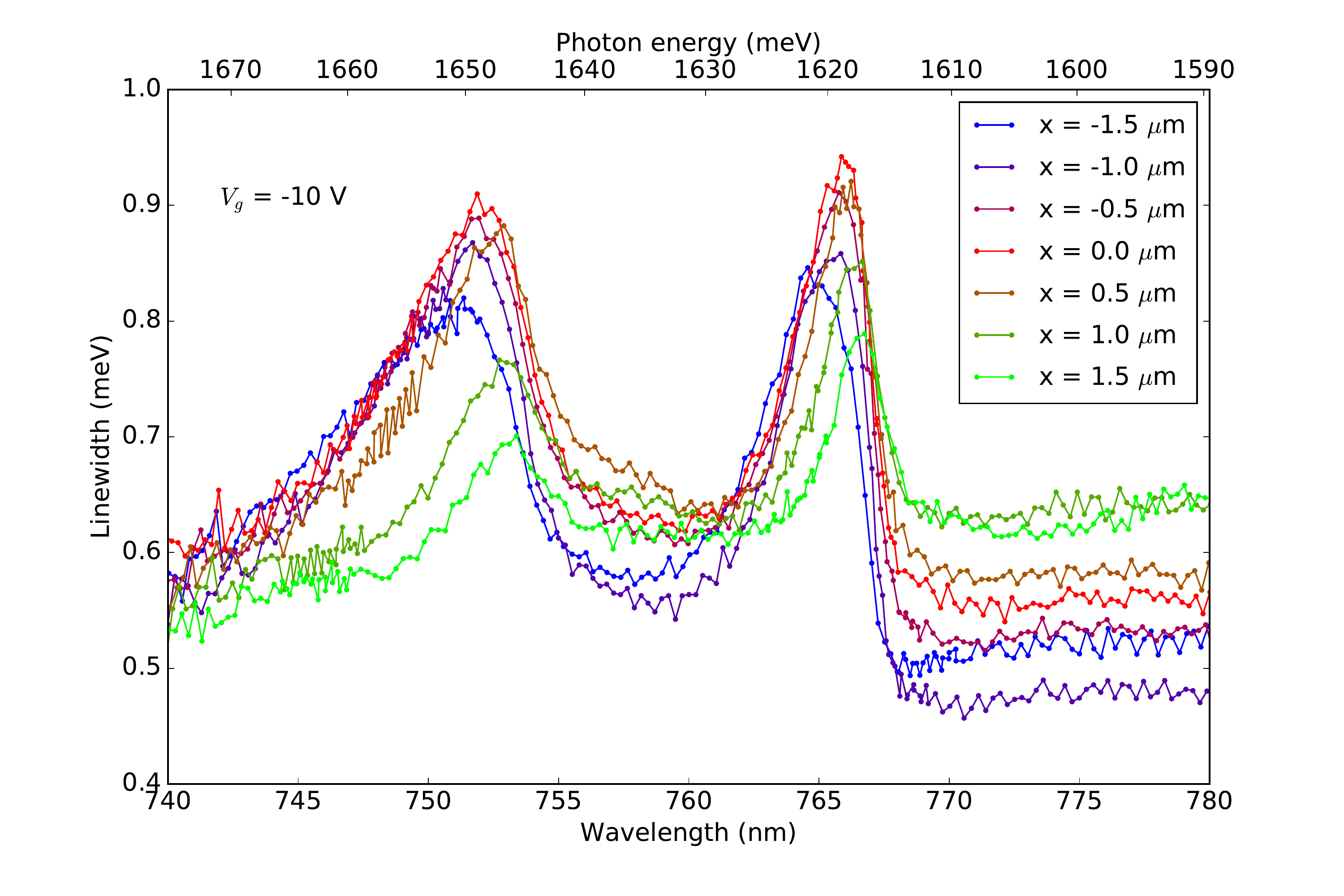}
		\caption{Spatial dependence of the absorption spectrum. }
\end{center}
\end{figure}

\section{Cavity mode fitting}

The source for the transmission spectroscopy is a LED  centered at
$\sim 760$~nm with a FWHM of $\sim 20$~nm. For Figure~2a,
Figure~3a,b,c and Figure~4a of the main text, the transmitted
spectrum is normalized by the LED spectrum. For the derivation of
the absorption spectrum of the flake from the transmission spectra
shown in Figure~2a of the main text, the normalization with the LED
spectrum is not necessary since only the linewidth and the center
wavelength of the cavity mode but not the intensity of the
transmitted cavity peaks are used to extract the absorption
spectrum. In addition to the fundamental cavity mode, higher
transverse modes are observed in the transmission spectrum. For
fitting the lorentzian peak to the transmitted fundamental cavity
mode, only the spectrum within a 4~nm wide window was considered in
order to exclude distortions from the higher transverse modes.

At the cavity lengths used for weak coupling measurements,  more
than one fundamental modes are observed within the stop band of the
DBR. For the derivation of the energy shift of the cavity due to the
resonances of the flake, the cavity length is calculated from the
wavelength of the next (lower-energy) fundamental mode. Since the
energy of that cavity mode is smaller than any resonances of the
MoSe$_2$ monolayer, its center wavelength serves as a good measure
for the cavity length.

We note that for the cavity length of $9.1\, \mu$m used to obtain the
data depicted in Fig.~2b, the exciton and the cavity are in the
strong coupling regime for $V_g < -10$~V. As a consequence, it is not
possible to obtain the imaginary part of the MoSe$_2$ linear
susceptibility by measuring the excess cavity line broadening as we
tune the cavity across the exciton (repulsive polaron) resonance. As
a remedy, we chose to plot the linewidth of the cavity-like
polariton peak for the parameter range corresponding to the dashed
box in Fig.~2b. The cavity line broadening we extract in this manner
is larger than the actual width of the exciton resonance. It does
however, yield the correct exciton resonance frequency. As an
alternative, it is possible to extract the actual imaginary part of
the MoSe$_2$ linear susceptibility by fitting the data to a formula
that describes the absorption lineshape in the presence of strong
coupling \cite{atac1990PRL} and extract an exciton linewidth of
4.5~meV. The drawback of the latter formula is that it is only
accurate if the electronic resonances are Lorentzian; this is
satisfied only for $V_g < -10$~V.

\end{document}